\documentclass[traditabstract]{jcil}
\usepackage[utf8]{inputenc}
\usepackage[round,colon]{natbib}
\usepackage{amssymb}
\usepackage{graphicx}
\title{Conspiratorial cosmology. II. The anthropogenic principle}
\author{J\"org P. Rachen\inst{1,2,3}\thanks{Email: universe23@jpr-cosmic.de}
\and Ute G. Gahlings\inst{2}}
\institute{\small 
Institut f\"ur analytische Zahlenmystik, Rautavistische Universit\"at Grafenhausen, Germany
\and Institut f\"ur angewandte Oligophrenie, Rautavistische Universit\"at Gr\"afinnenhausen, Germany
\and Astrophysical Institute/Interuniversity Institute of High Energy, Vrije Universiteit Brussel, Belgium}

\date{Received April 1, 2020; Accepted April 23, 2020; Published April 23, 2020.}

\begin{document}



\abstract{%
We revisit our 2013 claim that the Universe is the result of a conspiratorial plot, and find that it cannot be trusted, because even the belief in this conspiracy likely results from a conspiracy. On the basis of mathematical beauty, the final results of the Planck mission, the exploration of the dark sector by means of occult rituals and symbols, and a powerful new philosophical approach to physics, we demonstrate here that not only the existence of our Universe but the whole concept of reality has to be rejected as obsolete and  generally misleading. By introducing the new concept of the ``anthropogenic principle'', we eventually illuminate the darkest corners of the conspiracy behind the conspiracy and briefly discuss some important implications regarding the survival of wo*mankind.\thanks{%
We introduce here the \textit{gender asterisk} used in German to create a gender neutral language. It was recently elected \textit{Anglicism of the Year} although (because?) it has hardly any applications in English language -- here is at least one.}
}





\maketitle

\section{Introduction}\label{intro}

In a seminal paper based on the first cosmology results of the Planck Mission \citep{PlanckMission}, the authors provided compelling evidence that our Universe grew out of a conspiratorial plot \citep[][hereafter Paper\,I]{Consp23}, inferred from the discovery that the most important cosmological parameters can be derived  from the conspiratorial numbers $\pi$, $23$ and $42$  \citep{Archimedes, Illuminatus, Adams} by simple calculus. The paper received significant attention \citep[e.g.,][]{DiSia}\footnote{%
An excellent overview about this and its relation to other important work on the subject can be found on the following website:  https://www.soulask.com/?s=Conspiratorial+Cosmology}, but experts in the field \citep{SkyTelescope} kept questioning especially the role of the numbers $23$ and $42$: Are these really \textit{fundamental numbers of conspiracy theory}, or are they themselves \textit{product of a conspiracy} and only point to a more fundamental underlying truth that may not even be known to the deepest initiates?

The most fundamental principle of conspiracy theory (MFPCT) states that \textit{whenever you think it is something, it is for sure something else}. It follows that those who are least known to be inclined to conspiracy theory are the most suspicious ones, especially if They (see Paper\,I for notation)
have demonstrated ability to gain deep insight. One of the foremost individuals of this kind is the mathematician Leonhard Euler, who left us with the enigmatic identity \citep{MathAndImagination}
\begin{equation}
\label{EulerIdentity}
e^{i\pi} + 1 = 0\;.
\end{equation}
It has the property to make mathematicians feel like poets reading a Shakespearean sonnet \citep{Devlin}, because it connects the five most fundamental numbers of mathematics in the most simple way: the base of the natural logarithm, $e$, the imaginary unit, $i$, the circle-number, $\pi$, and the identities of multiplication, $1$, and addition, $0$.  Moreover, following Devlin:
\begin{quote}
\textit{[It] reaches down into the very depths of existence.}
\end{quote}
This is where we want to go, so we adopt the ansatz that these five numbers point to the base of all conspiracy. The easiest case is made for $\pi$, which we included as a conspiratorial number in the first place, and further evidence for its impact on cosmology has been delivered by \citet{PiInTheSky}. Moreover, $n=23$ is the only integer solution to the Scott-inequality (see Paper\,I) 
\begin{equation}
\label{ScottInequality}
\pi^e < n < e^\pi\;,  
\end{equation} 
which hides a link of conspiracy theory to the natural base $e$. As we have so far only considered a conspiratorial origin of reality, we postpone the discussion of the imaginary unit $i$ to Sect.\,\ref{enigma}. The most mysterious numbers in Eq.\,\ref{EulerIdentity} are $1$ and $0$: as all conspiratorial correspondence is based on multiplications, these numbers appear useless as they either do not change the result or annihilate it. But applying the MFPCT, it is this very paradox that supports us in believing that Eq.~\ref{EulerIdentity} is the crux of the matter, hence to the answer to the ultimate question: Who. Are. They?

\begin{table*}[t]
\begin{tabular}{lcccc}
\hline
\multicolumn{2}{c}{base parameter} & $\cong_{2013}$ & $\cong_{2015}$ & $\cong_{2018}$
\\\hline
\makebox[0.555\textwidth]{Physical baryon density\dotfill} & 
\makebox[0.15\textwidth]{$\omega_b \equiv \Omega_b h^2$\dotfill}
& $23{c}$ & $23{c}$ & $\phantom{|^2}{23{c}^*}^{\phantom{|^2}}$ \\ 
\makebox[0.555\textwidth]{Scaled physical matter density\dotfill} & 
\makebox[0.15\textwidth]{$\Omega_m h^3$\dotfill}
& ${c}$ & ${c}^*$ & ${c}$ \\ 
\makebox[0.555\textwidth]{Redefined acoustic scale measure\dotfill} & 
\makebox[0.15\textwidth]{$\theta_*' \equiv 100\theta_*{-}\,1$\dotfill}
& $42$ & $42{c}^*$ & $42{c}^*$ \\ 
\makebox[0.555\textwidth]{Thomson scattering optical depth due to
reionization\dotfill} & 
\makebox[0.15\textwidth]{$\tau$\dotfill}
& ${c}$ & $23\pi{c}$ & $23^2$ \\ 
\makebox[0.555\textwidth]{Scalar spectrum power law index \dotfill} & 
\makebox[0.15\textwidth]{$n_s$\dotfill}
& ${c}$ & ${c}$ & ${c}$ \\ 
\makebox[0.555\textwidth]{Log power of the primordial curvature
perturbations\dotfill} & 
\makebox[0.15\textwidth]{$\ln(10^{10}A_s)$\dotfill}
& $\pi$ & $\pi^*,\pi{c}^*$ & $\pi{c}$ \\ 
\makebox[0.555\textwidth]{Dark energy density divided by critical
density today\dotfill} & 
\makebox[0.15\textwidth]{$\Omega_\Lambda$\dotfill}
& $23\pi{c}$ & $23\pi{c}$ & $23\pi{c}^*$\\
\makebox[0.555\textwidth]{Matter density today decided by critical
density\dotfill} & 
\makebox[0.15\textwidth]{$\Omega_m$\dotfill}
& $\pi$ & $\pi$ & $\pi$\\ 
\makebox[0.555\textwidth]{Current expansion rate in km/s/Mpc\dotfill} & 
\makebox[0.15\textwidth]{$H_0$\dotfill}
& $23\pi{c}$ & $23\pi{c}^*$ & $23\pi{c}^2$\\ 
\makebox[0.555\textwidth]{Redshift at which the Universe is half
reionized\dotfill} & 
\makebox[0.15\textwidth]{$z_{\rm re}$\dotfill}
& ${c}^*$ & ${c}$ & $23\pi$\\  \hline
\end{tabular}
\caption{\label{parameters}\small Conspiratorial correspondence of Planck parameters as a function of Planck analysis cycles. For values marked with an asterisk the correspondence is within the standard conspiratorial confidence interval of 23 decisigma, otherwise it is 1 sigma.}
\end{table*}

One thing must be clear: We cannot expect that answering this question is a sure-fire success. Even if we have already found in Paper\,I that Conspirators are malicious but not subtle, we cannot assume the same for the Inner Circle that is \textit{really} behind everything. So it is no surprise that the recent attempt by one of the authors to link cosmology to astroparticle physics and chaos theory in order to relate the structure of the Universe to the communist world conspiracy \citep{Manifesto} has been reluctantly received by the community. In this paper, we shall get to the bottom of it and leave no stone unturned until we have illuminated all abysses with the bright glow of truth. So here we go.

\section{Planck 2015/18 results and superconspiratorial shifts in cosmological parameters\label{Planck}}

Since the publication of Paper\,I in the context of the first Planck data release, Planck has presented two further data releases: The \textit{Planck 2015 results} \citep{Planck2015Overview} that for the first time considered CMB polarization,  and the \textit{Planck 2018 results} \citep{Planck2018Legacy} that presented an extensive reanalysis and are expected to appear in A\&A within less than a Hubble time. Table \ref{parameters} gives an overview of the development of the conspiratorial correspondence of the parameters discussed in Paper\,I with the Planck analysis cycles. For some parameters we observe a striking consistency in their conspiratorial message (e.g., $\Omega_m \cong \pi$, $n_s\cong c$), for others we see the action of superconspiracy expressed by an additional factor $c$
($\theta'_*$, $\ln(10^{10}A_s)$), which has also been observed in the transition from WMAP \citep{WMAP} to Planck parameters (see Paper\,I).\footnote{%
We note again that $c = 23\cdot 42 = 966$ is the superconspiratorial constant, and we want to express our indignation that careless physicists, despite our haunting appeal in Paper I, do not stop to use the same symbol for the velocity of light.} 
But beyond this, there are two major \textit{new} discoveries which promise to open new gates for our quest: The first is the reionization optical depth parameter $\tau$ for which we found hints for superconspiracy in Paper\,I, but the large errors did not allow to make a final assessment. Consequently the Planck Collaboration has put a major effort in the more accurate determination of this parameter \citep{PlanckReionisation} and obtained an intriguing result: $\tau \cong 23^2$, a clear indication that there is a conspiracy behind the conspiracy! 

The second is the appearance of squared superconspiracy, $c^2$, in the Hubble parameter $H_0$. In Paper\,I we still excluded this possibility following the corollary that \textit{squared superconspiracy is imbecilely unstable}.\footnote{%
To proof this corollary, the interested reader is asked to choose any odd or obscure lines of argument from the literature to adapt them as needed. We recommend to close the harangue either with the phrase \textit{quod erat demonstrandum (q.e.d.)}, or by repeating the assumption after the words ``this proves that'', or something similar. As introductory literature we recommend \citet{Thomas}, who proved the existence of God, \citet{Nietsche} who proved the non-existence of God, and \citet{Heidegger} who proved the existence of the non-existence.} 
As it is observed now, we have to follow \citet{Kuhn} and perform a paradigm shift: \textit{Squared superconspiracy is possible in an environment of prevailing imbecility}, and as $c^2$ emerged in $H_0$ between the Planck 2015 and 2018 results, we assume that an \textit{inflationary release of imbecility} happened sometime in between.  But which event could this have been? 

It has been shown that (i) inflationary imbecility can be connected to the solution of unsolvable problems \citep{Asterix}; 
%
(ii) They always take some precautions to keep us away from Them; (iii) following the Chuck-Norris-theorem, there is always an easy way and a hard way \citep{Norris}. As we expect Them to be at least at the wisdom level of martial arts masters we assume that They will always give us a chance and try the easy way first: keep us busy!  So we suppose that They repeatedly supplied us with unsolvable problems, e.g., the geometrical squaring of the circle to cast a spell on mathematicians from antiquity \citep{quadrature} to modern times \citep{Lindemann}, or questions like how many angels fit on a needle-point to occupy monks in scholastic debates \citep[for a summary, see][]{Morgenstern}.  
Now we suspect that after the revelations of our 2013 work brought them into rough seas, They panicked so badly that They decided to keep all wo*mankind busy for some while and gave them the task: \textit{Elect and unelectable president!} How perplexed must They have been to see that hardly three years later the problem was solved \citep{USElections}! Although yet speculative, the enormous amount of inflationary imbecility released by this event likely pervaded the Planck data and caused the observed superconspiratorial shifts in the Planck 2018 cosmological parameters including the rise of squared superconspiracy.

But of course the story is not over: The authors are perfectly aware that They are perfectly aware that we (the authors) do not give up on pursuing them, so we suspect that by the time we decided to take the next step, They did as well: They followed an earlier suggestion of the initiate Dean R. \citet{Koontz} and presented wo*mankind the next unsolvable problem: \textit{Control an uncontrollable virus!} So we strongly suggest to the post-Planck community that when hopefully one day this pandemic is over and rationality has returned, to re-analyze the Planck data once more. Depending on how the dangerous brew of panic-brushed media, toilet paper stacking hysterics, obstinate scientists and erratic politics continues to boil, bubble and swirl in the cauldron of a real-life exponential function, we would not be surprised if even higher powers of $c$ are found, potentially even hyperconspiracy ($c^c$), which would point to the existence of the yet hypothetical \textit{dark unified mega-bunkum} (DUMB)---but this is beyond the scope of this paper.
  
\section{Meeting the dark side\label{dark}}

\subsection{Basics and experimental setup}\label{dark:exp}

One of the main revolutions of modern cosmology is the insight that most of the interesting stuff in the Universe is dark: dark energy, dark matter, dark ages. Only for the latter, ``dark'' has to be understood as the absence of light, so that it was possible to discover its conspiratorial content by scientific methods ($\tau\cong 23^2$ is the cosmological parameter which determines the duration of the dark-ages). For the former two, however, ``dark'' has a different meaning: it denotes the hidden, the obscure, and as a matter of fact all attempts to enter this ``dark sector'' with scientific methods have failed so far. We therefore have to check for alternatives, and find them in a meta-scientific approach that was rediscovered from ancient mythology in the 19th century by \citet{Levi}, and further developed by, among others, \citet{Blavatsky} and \citet{Crowley}: \textit{Occultism}. 

Occultism has a quite different approach to reality than regular science. The basic theory essentially consists of the spirits and mythical hybrid creatures, the methodology is performing rituals. Experiments usually involve a dark room, lots of candles, drugs cooked from certain mushrooms, herbs and fruit, and in most cases a medium, i.e., a person who can talk to the spirits.  Motivated by recent evidence for dark matter being a sphinx \cite[and references therein]{Mirabel}, we decided to set up an experiment in form of a séance to make contact with the dark sector. In the following we give a brief protocol of the experiment.

\begin{figure}
\centerline{\includegraphics[width=0.923\columnwidth]{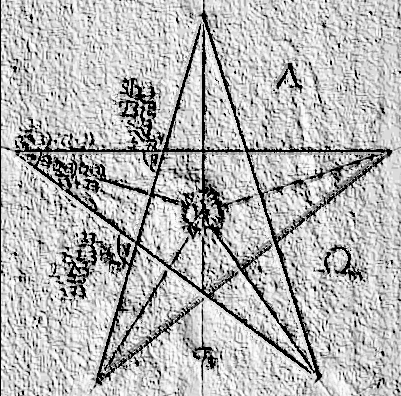}}
\caption{\label{pentagram}\small Greatly enlarged photograph of the pentagram that was burned into our wall plaster after our occult contact to the the dark sector (see text).}
\end{figure}

We sat with a group of people in a dark room illuminated by candles around a round table, with the famous medium \textit{Rettam Krad} in the midst of us. We had our hands on the table such that we were touching all our fingers left and right to close the magick circle, while the medium murmured some verses in a non-existent language. After a while, the candles flickered, the medium went into a trance and made strange noises, when suddenly a spot of extreme darkness, darker that anything we ever had seen, appeared in the middle of the table, and the medium screamed: 
\begin{quote}
\textit{Neutrino fertilis!}
\end{quote}
Then it smelled burned and a little like sulfur, and the lights came on. The medium collapsed. Then one of us pointed to a small, about 5mm large black dot in the wall plaster, and upon closer inspection we discovered that it was a pentagram surrounded by symbols, shown in Figure \ref{pentagram}.\footnote{%
We do not want to forget to draw the readers attention to the fact that the pentagram can have some very unpleasant features when it is pointed down \citep{Levi}. We therefore strongly advice the reader \textit{never} to hold this paper upside down!}
 
\subsection{Analysis and properties of the fertile neutrino}\label{nu:fert}

\textit{Neutrino fertilis}---the fertile neutrino---this is the clear message how to solve the dark sector problem. In the pentagram depicted in Figure \ref{pentagram} we see its symbol in the center, $\nu_f$, and arrows from all of its five corners point to it. We suspect that the five corners of the pentagram stand for the five numbers in the Euler-identity, Eq.~\ref{EulerIdentity}, although this requires further confirmation. Around the pentagram on the right side, we see the cosmological symbols $\Lambda$ and $\Omega_m$, the central parameters for dark energy and (mostly dark) matter. At the bottom, we see $\sigma_8$, the parameter which describes structure formation \citep[and maybe even ultra-high energy cosmic rays,][]{Manifesto}. The scribble at the left side of the pentagram will be discussed later.

How to put all this together was convened to the authors in a shared dream \citep[generally accepted as a method of scientific theory building, e.g.,][]{Kekule}: We saw a universe full incredibly happy and cheerful fertile neutrinos, which felt a bit lonely being equally distributed in space. They started singing ``Come Together'' \citep{Beatles} and danced towards each other, and whenever a considerable group of the funny little pals had gathered they started a spontaneous swinger party. And as fertile neutrinos are \textit{very} fertile, at least up to several hundred gpu (guinea pig units), they made many little cheerful neutrobambini, and in order to give them a place to live, they created for each of them a little bit of spacetime. And after the neutrobambini were quickly grown up they cheerfully joined in with the hustle and bustle, and so it continued.

The interpretation of this dream is clear: there is no such thing as dark energy, rather both accelerated expansion of space and contraction of matter goes back to the fertile neutrino, the main constituent of dark matter---or to say it in Newspeak \citep{Orwell}:
\begin{quote}
\textit{Expansion is contraction.\\ Contraction is expansion.}
\end{quote}
Obviously this solves several problems in modern cosmology, for example the question why accelerated expansion occurs just now when structure is getting established.  Unknown yet is whether it might explain the discrepancy between $H_0$ in the early and late Universe \citep{Planck2018Legacy, HubbleHubble}, and what its relation is to the much discussed \textit{sterile neutrino} \citep[for an introduction see][]{Naumov}. The main difference between these two hypothetical particles seems to be that the sterile neutrino is predicted but may be not observable, while the fertile neutrino is observed (we have seen it in our dreams) but may be not predictable.\footnote{Parallel to the release of our preprint \citet{CDE} proposed a form of dark matter with increasing mass from coupling with dark energy. Enlightened by our magical experience we recommend the authors to take a closer look at the procreation of neutrinos in order to avoid confusing cause and effect, and to become aware that the particle they are looking for is nothing else than the \textit{fertile neutrino}.}

\subsection{Conspiratorial connection}\label{dark:consp}

From the conspiratorial point of view, however, the most relevant part of the pentagram's message is hidden in the scribble on the left side, which shows repeatedly the numerals $2$ and $3$ in all kind of combination, and is apparently related to the angles in the pentagram of witch there are three: $36^\circ$, $72^\circ$, and $108^\circ$ (the unit degree is of no conspiratorial significance and ignored henceforth). First we note that all these numbers have a deep meaning in mythology and related areas: there are 36 chambers of the Shaolin, in Islam is 72 the measure for the pleasure of the martyr in paradise, there are 108 steps in the dance of Shiva, 108 disturbing emotions in Tibetian Buddhism, and 108 postures in the long Yang style Tai Chi Chuan form.  But also from the point of view of analytical number mystics these numbers are interesting, as we find that $36 = (2{\cdot}3){\cdot}(2{\cdot}3)$, $72 = 2^3{\cdot} 3^2$, and $108 = 2^2{\cdot}3^3$---and a careful inspection of Fig.\,\ref{pentagram} shows that it contains precisely this message on its left. If we put the numerals in the right order and ignore operation signs and other irrelevant stuff, we see in each of them two times the $23$: serial in the $36$, parallel in the $108$, and in the $72$ as well after applying mirror-symmetry of Baphomet-Toth manifolds (a very similar concept exists in string theory, see next section). This is the conspiratorial message we received from the dark sector: A triple hint to a conspiracy behind the conspiracy. We are on the right way.

\section{Methodological considerations\label{method}}

After being equipped with the essential weapons of such different areas as modern cosmology and occultism, we need to re-arrange our powers a bit before we strike our final blow, and discuss the relation between physics and philosophy.

\vspace{-4.2pt}
\subsection{Philosophy according to physics}\label{phil:phys}

While many physicists consider philosophy as some act of talking nonsense about physics during lunchtime (Enßlin, priv. comm.), there are occasions in which physicists get genuinely philosophical, true to the wisdom that all philosophy begins with wonder \citep{Plato}. Among the things which make physicists wonder is, for example, the sudden insight that theories in empirical science (it is generally accepted that physics is empirical science) contain parameters which cannot be derived from first principles but have to be obtained by experiment, and once measured, they take certain values. The wonder gets overwhelming when the physicists realize that if those parameters had values slightly different from what they actually have, there would be nobody in the Universe who could wonder. This shocking realization caused them to phrase the so-called \textit{anthropic principle} \citep[it is not known to the authors whether it was phrased during lunchtime]{Carter}, and like everything physicists propose it comes in weak and strong (cf.\ fundamental forces):  While the \textit{strong antropic principle (SAP)} essentially states that the big bang was nothing else than unlocking the delivery room for the divine birth of wo*mankind \citep{Tipler},$\!\!$\footnote{%
Above all, we know today that God was significantly longer in labor than originally proposed \citep[who reported one week]{Genesis}, i.e., $13.8$ billion years  \citep{Planck2018Legacy}. Nobody knows why She didn't ask for a Cesarean.}
the \textit{weak anthropic principle (WAP)} is acceptable even to childless atheists as it only states that the observed values of empirical parameters are as they are because of the very fact that they are observed, which could be paraphrased by:
\begin{quote}
\textit{We are in the Universe in which we are.}
\end{quote}    
Unfortunately physicists, in particular when they get philosophical, have the habit to continue asking why: \textit{Why are we in the Universe in which we are}? This leads obviously to an infinite loop, so in order to stop this infantile behavior They sent a team of super-nannies under the lead of Chief-Conspirator Edward Witten who brought the whining physicists the ultimate pacifier called \textit{string theory}, including the redesigned super-suckler, \textit{M-theory} \citep{M-Theory}.

String theory is very good. It can not only explain everything and predict anything, it also generates a sufficient number of universes (${\sim}10^{{\sim}100}$) for the WAP to be reduced to a simple statistical selection effect: The existence of wo*mankind is just the result of a very low chance probability applied to a very large number of trials, so the WAP becomes natural. Consequently, the combination of WAP plus string theory (MWAP, not to confuse with WMAP) could be paraphrased by
\begin{quote}
\textit{We happen to be in the Universe\\ in which we happen to be.}
\end{quote}
This fundamental insight could have allowed us to continue our sweet dreams of existence, had not \citet{Scott} provided us with the conjecture that
\begin{quote}
\textit{[\ldots] there are other universes out there in which string theory is not only simple and correct, but even falsifiable as well.}
\end{quote}
Employing the same argument as for MWAP, we can then follow that even if the probability $p_f$ for the Scott-conjecture to apply in a random universe is extremely low, the number of universes in which string theory is falsifiable can still be expected to be very large. Moreover, as the prior probability for a random theory to be wrong is significantly larger than the probability to be correct, we can conclude that in addition to the \textit{one} universe in which it is \textit{falsifiable and correct} there are many more universes in which string theory is actually \textit{falsified and thus wrong}. This of course would mean that not only those universes, but also all other universes (including our own) would have to vanish immediately as they are the result of a wrong prediction. It is clear that such instability does not comply with the conspiratorial requirement for argumentative power (CRAP)\footnote{%
For the non-initiated reader we note that the CRAP requires for all theories in the conspiratorial context to be inherently irrefutable as follows: \textit{Whenever something is brought forward to refute a conspiracy theory, it is identified as part of the conspiracy and thus confirms the theory.}}
so that we have to reject MWAP altogether. 

\subsection{Physics according to philosophy}\label{phys:phil}

Unlike physics, philosophy got past its plainly-wonder-phase several millennia ago and thus became able to devote itself to the more serious issue how the alleged reality which is called \textit{nature} is related to the alleged reality we call \textit{mind}. Starting with the attempt to compare mind with a kind of ink pad \citep{Locke} a philosophical current called \textit{empiricism} was founded, and rediscovered in the early 20th century by the amazing realization that \citep{Wittgenstein}: 
\begin{quote}
\textit{The world is everything that is the case.}
\end{quote} 
Based on this, the initiate Rudolf Carnap, member of the influential lodge called the \textit{Vienna Circle} \citep[e.g.,][]{Stadler}, reduced the minds of scientists essentially to a data-processing machine constructing an exact reflection of reality just by logical operations on data \citep{Carnap1}. From the conspiratorial point of view, this is of course what They want us to think, so in spite of some problems to explain how abstract entities of modern physics (like ``black holes'', ``elementary particles'', and so on)  could possibly arise from data-processing  \citep{Carnap2}, They successfully tranquilized the science community with the lullaby that one could not even question logical empiricism---for further progress in this direction, see \citet{Ensslin}.

It should be noted, however, that there was another stream in philosophy of science called \textit{rationalism}, which was founded on the bold statement by \cite{Descartes}:
\begin{quote}
\textit{Cogito ergo sum.}
\end{quote}
(I think, so I am).\footnote{%
Actually Descartes wrote this first down in French, but later thought that good quotes sound significantly smarter in Latin. We agree to this, and will henceforth phrase central wisdoms of our teaching in Latin. Unfortunately none of the authors is an eminent authority in Latin, so we encourage all Latin teachers bored by their (non)-existence to send corrections to the email address given in the paper head---we will write all of them a hundred times.}
Rationalism opposed empiricism by means of the provocative idea that scientists may actively use their minds when doing science. It was insinuated to scientists that they develop ideas about nature, which they call hypotheses, theories or models, and confront them with experimental data in order to confirm or falsify them. This way, scientific progress would be described as a random walk governed by trial and error \citep{Popper}, dogma and revolution \citep{Kuhn}, administrative diplomacy \citep{Lakatos}, or pure anarchy \citep{Feyerabend}. Regardless of these details, it is obvious that the idea that imagination may be somehow related to what we call reality might have got Them in trouble if they had not stopped this movement. So They chose the hard way and sent a Cleaner \cite[e.g.,][]{Besson, Tarantino}, this time in person of the ``common sense philosopher'' David Stove, who recommended himself for the job by various socio-cultural articles \citep{Stove1} in the spirit of his great masterminds \citep{SpanishInquisition,Hexenhammer}. He identified Popper, Kuhn, Lakatos and Feyerabend as ``modern irrationalists'' \citep{Stove3},$\!\!$\footnote{%
The inclined reader may have noted that Stove did the job before he applied for it, but why should They who make the Universe care about causality?}
i.e., as cognitive heretics against the holy induction, and declared them outlaws by edict. Eventually they were burned at the stake of logic by E.\,T. \citet[Part I Sect.\,9.16.1]{Jaynes}\footnote{%
Some readers of Jaynes' otherwise excellent book might not have expected to find such an irritating section there, but this should not be regarded a surprise as \textit{nobody expects the Spanish Inquisition} \citep{MontyPython}.} 
on his crusade for the true faith.\footnote{%
We refer here to the religious war between the reformatory Church of Reverend Bayes (a) and the orthodox Church of Frequentism (b), which stemmed from the scholastic debate whether science is about (a) asking relevant questions to which there is no precise answer, or (b) giving precise answers to irrelevant questions (Desch, priv. comm.).}

\subsection{The exceptionicistic revolution}\label{exception}

As our analysis of the classical relation of physics and philosophy has not brought us forward in our quest, we have to turn to more recent developments. The most promising for our goals is hereby \textit{exceptionicism}, which is an extension of confirmation theory by the common sense principle:
\begin{quote}
\textit{Exceptions confirm the rule.}
\end{quote} 
Here, a ``rule'' has to be understood as a test implication related to theories or models \citep{Popper}.  Exceptionicism now states that \textit{the most trustworthy rules} (resp. the theories or models they stem from) \textit{are those which are exclusively confirmed by exceptions}. While logic and philosophy are still hesitant to accept this proposition, scientists from all areas have already left an impressive trail of exceptionicistic reasoning, let it be in gender-related psychology \citep{Moebius,Weininger}, about Milanković-cycles \citep{Milankovic} as an explanation for Earth climate \citep[e.g.][and many more]{Zachos, Wunsch}, astrophysics of radio sources \citep{MarscherGear, MPIfR, PlanckRadioSources}, or on the origin of ultra-high energy cosmic rays \citep{AugerStarburst}. And as everything proposed by physicists (mainly), also exceptionicism comes in weak and strong: \textit{weak exceptionicism (WEX)} allows that the models occasionally happen to fit the data, while \textit{strong exceptionicism (SEX)} demands immediate rejection of a model if it has any correspondence to experiment or observation. Although SEX is very appealing (not only) from the conspiratorial point of view and may in fact turn out the right methodology to deal with the fertile neutrino, we consider it premature to apply it in general, in particular as most of the successful examples of exceptionicism mentioned above would have to be rejected because occasional correspondence to reality is hard to avoid.

It is undeniable that exceptionicism is a very important tool in our search for the conspiratorial grail, but how can we best use it? Obviously, our methodology needs to fulfill the following desiderata: (a) It must be tolerant in terms of academic standards; (b) it must include the teachings from the dark side; (c) it must comply with the CRAP; and (d) it must kick Their butts.  Out of the many different philosophical approaches, it is clear that only Paul Feyerabend's view of science as \textit{an essentially anarchic enterprise} with its slogan \textit{anything goes}\footnote{%
Most physicists and some philosophers consider this statement one of the most ridiculous ever made in science history, which we believe is only thanks to the fact the Feyerabend failed to phrase it in Latin.}
\citep{Feyerabend} serves our needs in all points. The only weakness is that even the anarchic theorists are supposed to feel at least some discomfort when their theories do not correspond to the data. This can be cured, of course, by combining it with weak exceptionicism, which essentially states that it doesn't matter whether data are represented or not. And finally we hear from the off the Master's voice, the last great magician, Aleister \citet{Crowley} with his call to always follow your free will in love in order to succeed, phrased in a magick language. This combined we will then declare to our methodology and call it \textit{Anarchic Imaginism (AI)} along the lines of
\begin{quote}
\textit{Anything goes.\\
No matter what is measured or observed.\\
Do what thou wilt.}
\end{quote}
We call this the \textit{AND-principle}. Now we are ready to strike.

\section{The anthropogenic principle}

\subsection{Simulators and simulations}\label{sims}

There is no doubt that modern science would not be possible without computer simulations. We simulate everything: complex systems and non-complex systems, our perception and our conclusions, and if we do not understand the result of a simulation we run another simulation to do so. We simulate even problems which could be solved with a simple differential equation,\footnote{
We actually suspect that nowadays journals do not even consider papers that did not burn at least a million CPU hours.
} 
and it is foreseeable that soon elementary school kids will solve their math exercises (like $5 + \Box = 12$) by running simulations on their smartphones. So we have no doubt that almost every current PhD student and postdoc working in any area of science will immediately endorse the Cartesian paraphrase:
\begin{quote}
\textit{Simulo, ergo sum}
\end{quote}
(I simulate, so I am). It is worth to note, however, that before this outsourcing of intelligence to silicon-based structures began, there were people who considered the possibility of simulations running in a carbon-based computer,$\!\!$\footnote{%
As in particular younger scientists may never have heard of it: This carbon-based computer is called \textit{brain}. It consists neither of CPUs nor GPUs, it is somewhat like a neural network you know, just that (a) it has a much higher complexity and capacity, and (b) we understand even less what is going on inside it.}
realized that also this can be programmed, e.g., by the use of psychedelic drugs \citep{Lilly,Leary77},$\!$\footnote{%
We note here that these people the had a strong connection to the messengers of the conspiracy number 23 \citep{Leary83}.}
and concluded that what we call reality, including ourselves, is nothing but such a carbon-based simulation, or in Cartesian phrasing:
\begin{quote}
\textit{Simulor, ergo sum}
\end{quote}
(I am simulated, so I am). But what does this mean?

In Paper\,I we concluded from the evidence for cosmological conspiracy that our Universe does not exist, and is likely to be a computer simulation as described by \citet{Galouye}. Many readers may then have thought that this means we exist in some buzzing box in a computer center of a higher intelligence. \citet{Adams}, however, also discussed the opposite scenario, i.e., that we are just the processing units of a giant computer called Earth which was built to find the question to the answer 42. And finally, nobody doubts that \textit{AND}-gates exist in every computer and \textit{AI} plays a more and more important role in computing. So, could it be that Universe, including ourselves, is a simulation running inside ourselves? We may phrase this conjecture as
\begin{quote}
\textit{Ipsos simulamus, ergo universum est}
\end{quote}
(We simulate ourselves, so is the Universe), and of course in Latin, as we want it to sound smart. 

\subsection{Deciphering the Euler enigma}\label{enigma}

We return to the Euler identity, Eq.~\ref{EulerIdentity}, and try to assign meanings to the numbers in it beyond what mathematicians tell us (what do they know, anyway?). So let's start with the obvious: $e$, the ``natural base'', cannot stand for anything else than for \textit{nature}. Similarly, the ``imaginary unit'', $i$, stands obviously for \textit{imagination}. Not quite as simple, but still straightforward is the meaning of $\pi$, as it is related to circles, which in all cultures have been standing for the divine---remember that astronomers before Kepler believed that planets must move on circles because the considered them to be in the heavens. As the divine, God, also stands for creation, $e^{i\pi}$ may be read as ``imagination creates nature'', and also as ``nature creates imagination''. This obviously stands for the idea of ``self-simulation'', but who is running it?

Return to conspiracy theory: The number 23 is born out of the fundamental numbers $\pi$ (God) and $e$ (nature) via the Scott-inequality, Eq.~\ref{ScottInequality}, and it has lead us our way to arrive here. It appears, squared, in the cosmological parameter $\tau$ which determines the end of the dark ages, and also in trinity in the ancient magic symbol of the pentagram. So what is 42 then? We remember from Paper\,I that 
\begin{equation}\label{LogTrinity}
42_{10} = 101010_{2}\;,
\end{equation}
and noting that $2$ is the ``logical base'', we can read this as \textit{42 is the logical trinity of 10 in human-readable form}, because $10$ is the both symbol and base for the ten-fingered ape, $\alpha\nu\theta\rho\omega\pi o\varsigma$, {\textit{homo sapiens}, or simply wo*mankind. And if we recall the teaching of the dark side, the trinity of $23^2$ and its relation to the angles in the pentagram, we know that conspiratorial numbers can be hidden across mathematical operations, and looking at Eq.~\ref{EulerIdentity}, we see the number $10$ standing there across the equal sign. So this is it: The Euler identity is our conjecture from the end of Sect.~\ref{sims} phrased in mathematics, and this does not only sound even smarter than Latin, it also means that our conjecture is proven as Eq.~\ref{EulerIdentity}  is a mathematical truth. So in conspiratorial Newspeak this can be said as: 
\begin{quote}
\textit{They are Us.\\
We are Them.}
\end{quote}
This is the \textit{anthropogenic principle}, the realization that We\footnote{%
We see no reason to abandon Our notational conventions from Paper\,I}
are the ones who make everything: Life, the Universe and all the rest. And once We understand this, We can choose to sit in front of a white wall and listen for the clap of one hand \citep{Dogen}, We can kill the Creator and take over Her place \citep{Nietsche},$\!\!$\footnote{%
And don't forget the whip, Fritze!}
We can follow the white rabbit \citep{Carrol} and give the childlike empress a new name \citep{Ende}, or if We prefer, We can have dinner with ourselves in a Victorian hotel room at the edge of the Galaxy \citep{Kubrick}, unless We are busy with dropping books out of Our daughters'  bookshelf while sitting in the center of a black hole \citep{Interstellar}.  

\subsection{Practical applications}\label{survival}

The survival of wo*mankind is threatened---but not seriously, as we have demonstrated so many successful strategies to avert the danger.  The  symptomatically best formulation for those is found in Cologne dialect: \textit{Et hätt noch immer joot jejange} (engl: It has always gone well so far), and the best proof that it works is that, in spite of Carnival, Cologne still exists. Having such strategies, we do not need to worry about nuclear overkill, climate change, or any pandemics that may ever strike us. There is, however, a challenge coming up where all those strategies may not work any more, and if we do not want to fall into melancholic despair \citep{Krauss}, we need to be ready to fight the ultimate battle between Good and Evil: Life vs. Dark Energy \citep{Hooper}.

Space. The final frontier. These are the adventures of wo*mankind in the year 100 billion, who will have developed to a civilization of type III on the Kardashev scale \citep{Kardashev}, and there will be no place to boldly go any more as we will have been everywhere in the Galaxy already. The energy consumption of wo*mankind will have increased to $10^{36}$ Watts,$\!$\footnote{%
This estimate is made by taking the data of the world-energy consumption in the last 50 years and extrapolate them exponentially over the next few million years  -- a solid method of scientific prediction which cannot possibly be wrong.} 
which we need to harvest from all stars of the Galaxy via Dyson-spheres \citep{dysonsphere}. How all this energy will be used is less clear, because even if we assume that future wo*mankind will have spread over the Galaxy and inhabit 100 million putatively habitable planets and the energy harvested will be distributed over then, and considering that regardless of how this energy is initially used it is inevitably thermalized, all those planets would simply melt. But that should not be our concern here.

The challenge is now that in about 100 billion years the expansion caused by dark energy will take over the gravitational bound of the Galaxy, and our nice new big home will dissolve into emptiness. Depending on how much our intelligence has evolved over this time, there are then three options to react:

\noindent\makebox[2.3em]{\textbf{(i)}} If our intelligence is still at the same level as today, we will follow \citet{Hooper} and start pushing all our stars to keep the Galaxy together and our energy source accessible. In order to make sure that we need all the energy we harvest, we will build our spaceships in form of giant SUVs which are constructed such that they have a decent wind-resistance even in the interstellar medium, and we will cause a greenhouse effect in the Galaxy so strong that Dyson-trees \citep{dysontree} start to grow on molecular clouds. Eventually, the transdimensional creature Q \citep{StarTrek} will appear and give us the gold medal for the most stupid waste of energy in the Multiverse.  

\noindent\makebox[2.3em]{\textbf{(ii)}} If our intelligence will be advanced, but not yet to the point that we have reached satori \citep{Dogen}, we will use a few moments in the billions of years we have to think why we actually need all this energy, realize that we don't, and return to possibly a 100 million Kardashev I civilizations who, each on their own planet, can sit back, relax, and enjoy their increasingly dark night skies.

\noindent\makebox[2.3em]{\textbf{(iii)}} If we have reached satori and the anthropogenic principle is understood, we will call our best physicists, philosophers, magicians and other experts together, provide them with enough computers, food, wine and drugs to calculate us a new alleged reality in which the problem is avoided. We then restart the (self-)simulation with new parameters, and---besides some possible side effects, see \citet{Adams} for a selection---as long we made sure that some 100 billion years have to pass until something serious happens again, we will have time enough to sit back and relax.

\section{Conclusions}

So We have reached our goal: We found the conspiracy behind the conspiracy, and that is that We ourselves are the Conspirators. But what about the Universe now: does it exist or not? Our answer is: both or neither! And if both or neither \textit{a proposition and its negation} is true, logical philosophy teaches us that then at least one predicate in the proposition is meaningless. And in this case, this is the term ``reality'' -- it is a concept which simply makes no sense. Or to phrase it in Latin:
\begin{quote}
\textit{Realitas non datur,}
\end{quote}
and although We have to admit that similar thoughts have been thought before \citep[e.g.,][]{Buddha, Zeno}, We can now state to have proven it---on the solid base of mathematics and philosophy, guided by conspiracy theory and modern cosmology!

With this, all answers should be given---but have also all questions been asked? There are some questions we could imagine readers might ask us: Aren't you constantly contradicting yourself?---Yes, we do! Does anything you tell us make any sense?---No, it doesn't! Isn't it just a whole bunch of hooey you are telling us here?---Hey, you got it! And is this, as your last paper, not just an April Fool's joke?---Well\ldots\ let us now ask some questions to our readers: Do you really think that authors like Douglas Adams did not provide us with messages of deep truth just because of hiding them in funny books? Shouldn't we consider what the poet teaches us \citep{Horaz} 
\begin{quote}
\textit{Ridentem dicere verum / quid vetat?}
\end{quote}
(what prevents us from telling the truth with a laughter?)  No, although many of you may now sit back with a broad smile, we perfectly know that some our readers can't be fooled. They know how a apply the MFPCT, and that there is always something going on behind the scenes! And so the struggle continues. 

\vspace{-4.2pt}
\begin{acknowledgements}
We thank the scientific community for several decades of hostil\ldots no, hospitality. Good bye, take care, and have a nice day. From now on our only guideline is the omnipotent Dada.
\end{acknowledgements}


\vspace{-17pt}
\raggedright

\bibliography{}

\begin{thebibliography}{48}
\bibitem[{{Anaxagoras et al.}({470BC seq.})}]{quadrature}
Anaxagoras. fl.~470BC. Discovered the squaring-the-circle problem when killing time in prison; Eudemus of Rhodes. fl.~320BC. Gave the first argument why it might be unsolvable.
\bibitem[{{Archimedes}({fl.~250BC})}]{Archimedes}
Archimedes of Syracuse. {fl.~250BC}. Is considered the discoverer of the circle-number $\pi$.
\end{thebibliography}


\end{document}